\documentclass[prl,twocolumn,showpacs]{revtex4}
\def\journal #1, #2, #3, 1#4#5#6{{\sl #1~}{\bf #2}, #3 (1#4#5#6) }

\usepackage{amsmath}
\usepackage{mathrsfs}
\usepackage{float}
\usepackage{graphicx}
\usepackage{amssymb,amsmath}
\usepackage{indentfirst}

\def\eqa{\begin{eqnarray}}
\def\eea{\end{eqnarray}}
\newcommand{\eq}{\begin{equation}}
\newcommand{\ee}{\end{equation}}
\newcommand{\nn}{\nonumber\\}

\newcommand{\ua}{\uparrow}
\newcommand{\da}{\downarrow}

\begin{document}
%\begin{CJK*}{GBK}{song}

\title{Magnetism of Cold Fermionic Atoms on p-Band
of an Optical Lattice}
\author{Lei Wang$^1$, Xi Dai$^1$, Shu Chen$^1$,  and X. C. Xie$^{2,1}$}
\affiliation{$^1$Beijing National Lab for Condensed Matter Physics
and Institute of Physics, Chinese Academy of Sciences, Beijing
100080, China \\
$^2$Department of Physics, Oklahoma State University, Stillwater,
Oklahoma 74078, USA}

\begin{abstract}
We carry out \textit{ab initio} study of ground state phase diagram
of spin-1/2 cold fermionic atoms within two-fold degenerate $p$-band
of an anisotropic optical lattice. Using the Gutzwiller variational
approach, we show that a robust FM phase exists for a vast range of
band fillings and interacting strengths. The ground state crosses
over from spin density wave state to spin-1 Neel state at half
filling. Additional harmonic trap will induce spatial separation of
varies phases. We also discuss several relevant observable
consequences and detection methods. Experimental test of the results
reported here may shed some light on the long-standing issue of
itinerant ferromagnetism.

\end{abstract}

 \maketitle
%\end{CJK*}

{\it Introduction:} Recently, research of ultracold atomic gases
have simulated a new wave of studying the many-body problems. One
can create periodic potentials to confine the ultracold atoms by
intersecting laser beams. Because of the experimental
controllability of atom number, dimensionality, geometry as well as
interaction strength {\it etc.}, the optical lattice provides an
ideal playground for investigating many body problems. Specifically,
the experimental realization of a Mott-insulator phase transition in
the optical lattice \cite{Bruder}\cite{Greiner} brings us to the
forefront of the strongly correlated systems. More recently, the
experimental progress makes it possible to put fermions into the
optical lattices and control the interaction between them
\cite{Kohl}\cite{mi_fermion}. These advances open a new channel to
investigate numerous phenomena that play important roles in the
condensed matter physics, such as quantum magnetism
\cite{georges}\cite{duan}\cite{rvb}, high-temperature
superconductivity\cite{Hofstetter} and physic associated with band
degeneracy \cite{tlho}\cite{afho}.

Itinerant ferromagnetism in transition metals is another
controversial issue in condensed matter physics. Despite long
history dating back to Stoner \cite{stoner}, a fully understanding
of its mechanism is still lacking \cite{dmft_fm}. However, a
consensus is reached that the appearance of robust ferromagnetism is
not a generic feature of the single band Hubbard model in a cubic
lattice. \textit{Non-bipartite} lattice structures may stabilize FM
order, degenerate bands and the Hund' rule coupling may also help.
Testing various scenarios using the ultra-cold fermionic atoms in
optical lattice may shed light on this issue. The highly tunable and
neatness of cold atom experiments also permit calculations with no
adjustable parameters, the phase diagram and experimental signatures
could be determined through the following first principle
parameters: the depth of the optical lattice $\bf {V}$ for each
directions and the s-wave scattering length $a_s$.

In this paper, we first perform band structure calculations to
identify a parameter region of $\bf {V}$ and $a_s$ where fermionic
atoms on an optical lattice could be effectively described by a
two-band Hubbard model. Band calculation also provides effective
coupling strengthes of the model. We then obtain zero temperature
phase diagram as function of the coupling strength and the density
of fermionic atoms by applying
%Kotliar-Ruckenstein slave boson method.
Gutzwiller variational approach. Finally, we consider the effect
of external harmonic trap within the local spin density
approximation (LSDA), which is shown to induce concentric shell
structures of different phases. In particular, a robust
ferromagnetic shell composed of spin polarized fermions emerges.

\begin{figure}
\includegraphics[width=7cm, bb=16 17 310 219]{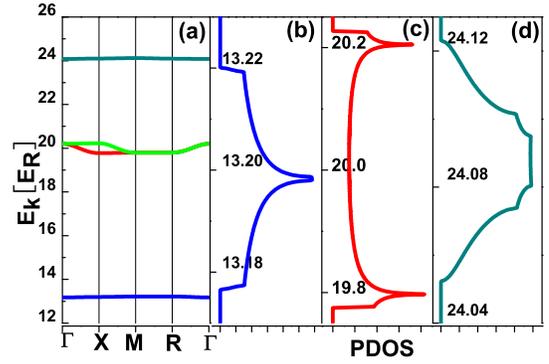}
\caption{(a). Band structures of four low lying bands ($s$, $p_x$,
$p_y$ and $p_z$) of an optical lattice with $V_x=V_y=16 E_R, V_z=36
E_R$. (b)-(d). Projected density of states (PDOS) for each band. The
bandwidth of $p_x$ and $p_y$ band (c) is about $10$ times wider than
the $s$-band (b), due to the large overlap of p-band atomic wave
function. The low lying $s$-band shows 2D characterizes, while PDOS
of $p_x$ and $p_y$ band are degenerate and show quasi-1 dimension
characterizes. A weakly breaking of the particle-hole symmetry also
could be recognized from the asymmetry of the PDOS. Interestingly,
the PDOS of $p_z$-band has 3D characterizes (d), this is due to the
strength of $\sigma$-bond along z-axis is strongly suppressed by
large $V_z$ and becomes comparable to the weak $\pi$-bond along $x$
and $y$ directions.
%, thus the hopping amplitude associate
%with vertical hopping is reduced to the same order of the weakly
%$\pi$ bond hopping within the plane.
\label{pdos}}
\end{figure}

{\it The model and the band structure calculations:} Six
counter-propagating lasers generate potential of the form
$V(\textbf{r})=V_x\sin^2(k_Lx)+V_y\sin^2(k_Ly)+V_z\sin^2(k_Lz)$.
Atoms trapped by this potential form a simple cubic lattice with
lattice spacing $a=\frac{\pi}{k_L}$. Fig.\ref{pdos} shows the band
structures of the optical latices with $V_x=V_y=16 E_R, V_z=36 E_R$,
where $E_R=\frac{\hbar^2k_L^2}{2m}$ is the recoil energy. The
degenerate $p_x$ and $p_y$-band are well separated from the low
lying $s$-band,  $p_z$ band are pushed up by the large $V_z$.
Assuming the $s$-band is fully occupied, as long as the band gaps
are much larger than the interaction scales within each band, we can
focus on the degenerate $p_x$ and $p_y$ bands. Authors of
\cite{wuzhao} assume that atoms mainly interact in the p-wave
channel thus could be effectively treat as spinless fermions and
they studied effects of different lattice structures. In the present
study, we focus on cubic lattice and consider the s-wave
pseudo-potential
$V(\bf{r}-\bf{r}^\prime)=\frac{4\pi\hbar^2a_s}{M}\delta(\bf{r}-\bf{r}^\prime)$,
which makes the problem more relevant to condensed matter system
where spin $1/2$ fermions plays the main role. Neglect all
interaction terms except the largest on-site one, we introduce an
effective two-band Hubbbard model on cubic lattices:

 \eqa H &= & H_{kin}+H_{int} \nn H_{kin} &=&
\sum_{\textbf{k},\alpha,\sigma }\ \varepsilon_{\textbf{k}\alpha}\
\hat{c}^\dagger_{\textbf{k}\alpha\sigma}\
\hat{c}_{\textbf{k}\alpha\sigma}+h.c.\nn
% H_{int}&=&  U\sum_{i
%\alpha } \hat{n}_{i\alpha
%\ua}\hat{n}_{i\alpha\da}+U^\prime\sum_{i\sigma\sigma^\prime}
%\hat{n}_{ix\sigma}\hat{n}_{iy\sigma^\prime} \nn &&- J\sum_{i}
%\hat{n}_{x\sigma}\hat{n}_{y\sigma} +\sum_{i \sigma} J
%c^\dagger_{x\sigma}c^\dagger_{y\bar{\sigma}}c_{x\bar{\sigma}}c_{y\sigma}
%\nn && +J\sum_{i}(c^\dagger_{x\ua}c^\dagger_{x\da}c_{y\da}c_{y\ua} +
%c^\dagger_{y\ua}c^\dagger_{y\da}c_{x\da}c_{x\ua})
 H_{int}&=&  U\sum_{i
\alpha } \hat{n}_{i\alpha
\ua}\hat{n}_{i\alpha\da}+U^\prime\sum_{i\sigma}
\hat{n}_{ix\sigma}\hat{n}_{iy\bar{\sigma}} \nn &
+&J\sum_{i\alpha}(\hat{c}^\dagger_{i\alpha\ua}\hat{c}^\dagger_{i\bar{\alpha}\da}\hat{c}_{i\alpha\da}\hat{c}_{i\bar{\alpha}\ua}
+
\hat{c}^\dagger_{i\alpha\ua}\hat{c}^\dagger_{i\alpha\da}\hat{c}_{i\bar{\alpha}\da}\hat{c}_{i\bar{\alpha}\ua})
\label{onsite}\eea

With \eqa U&=&\frac{4\pi\hbar^2 a_s}{M}\int d\textbf{r}
w_{x}(\textbf{r})^4 \nn U^\prime&=&J=\frac{4\pi\hbar^2 a_s}{M}\int
d\textbf{r} w_{x}(\textbf{r})^2w_{y}(\textbf{r})^2 \label{UandJ}\eea
where $\alpha=x,y$ is the band index, $\sigma=\ua,\da$ denote spins,
$\varepsilon_{\bf{k} \alpha}$ is energy dispersion for each band.
$a_s$ is the s-wave scattering length which could be tuned by
Feshbach resonance technique over a large scope of magnitude. We
only consider repulsive interaction ($a_s>0$) in this study.
Parameter $U$ describes the on-site Coulomb repulsive interaction
between atoms reside in the same orbital, while $U^\prime$ describes
the repulsive interaction for two atoms in different orbitals. $J$
controls the strength of Hund's rule coupling between different
orbitals. In the present system only transverse part of the Hund's
rule coupling survives, {\it i.e.} the spin flip and pair hopping
processes. The absence of the density-density interaction for atoms
with the same spins is due to the short range $\delta$ type
pseudo-potential. This fact can also be understood as a fully
canceling between the longitudinal part of the Hund's rule coupling
and corresponding terms in the inter-orbital density-density
interactions. All of the interaction parameters are expressed as
overlaps of maximally localized Wannier functions\cite{mlwf}.

Both the band dispersions $\varepsilon_{\bf{k} \alpha}$ and the
coupling constants in Eq. \ref{UandJ} are determined by
experimentally controllable parameters $\bf{V}$ and $a_s$. Due to
the anisotropic nature of the p-orbitals, the system shows dimension
reduction behavior: the PDOS has Van Hove singularity near the band
edge resembling of 1D character. This feature is also included by
the tight binding model with an anisotropic hopping amplitude along
different directions\cite{cjwu}. At half filling, the Fermi surface
has nesting property with the nesting vector $(\pi, \pi,
\pi)$\cite{zhai}. In contrast to the isotropic case (s-band), here
the singularity in PDOS and Fermi surface nesting occur at separable
energy scales and lead to remarkable consequence: instability
towards to different magnetic orders dominates at different
fillings. For the given lattice depth $V_x=V_y=16 E_R, V_z=36 E_R$,
our calculation shows the bandwidth $W$ of the two-fold degenerate
$p$-band is $0.45E_R$, while band gaps separated $p_x$, $p_y$-bands
with $s$ and $p_z$-bands are $6.79E_R$ and $4.10E_R$
(Fig.\ref{pdos}(b)). Overlap of Wannier functions gives
$U/E_R=8.63k_La_s$ and $U^\prime=J=0.299U$, while the interaction
scale within s-band is $13.1(k_La_s)E_R$. Not exceeding the band
gaps pose $0.0755\lambda$ as the upper limit for $a_s$.  Within this
restriction, one could still tune the system from weak to strong
coupling ($U\sim9W$) region by Freshbach resonance technique.

\begin{figure}
\includegraphics[width=4.1cm, bb=12 11 244 277]{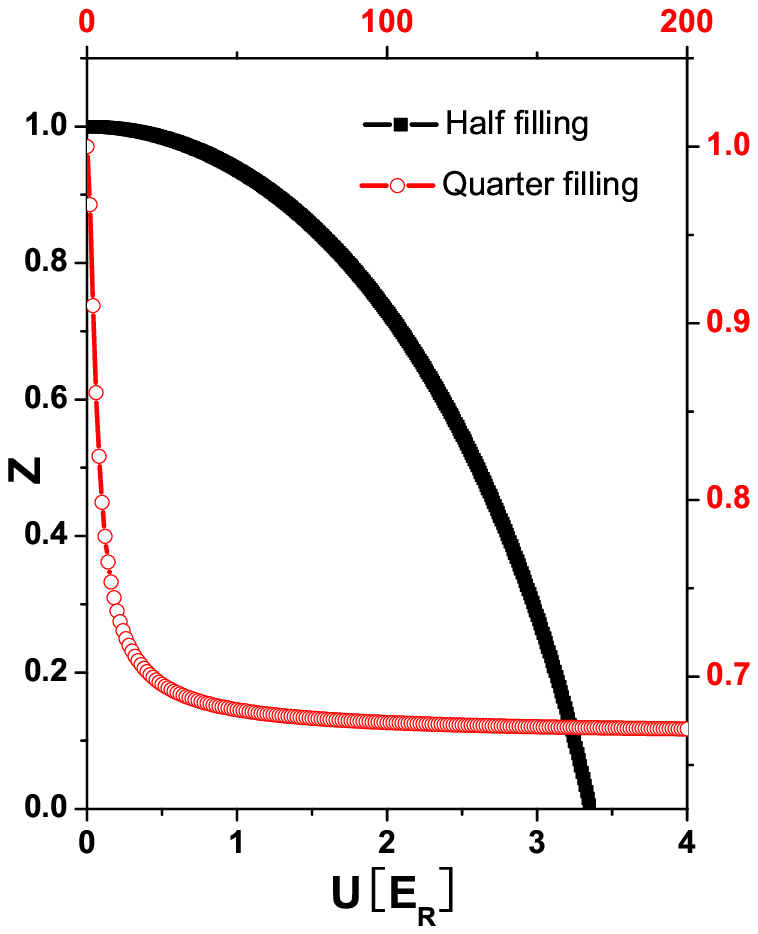}
\includegraphics[width=3.9cm, bb=10 16 216 257]{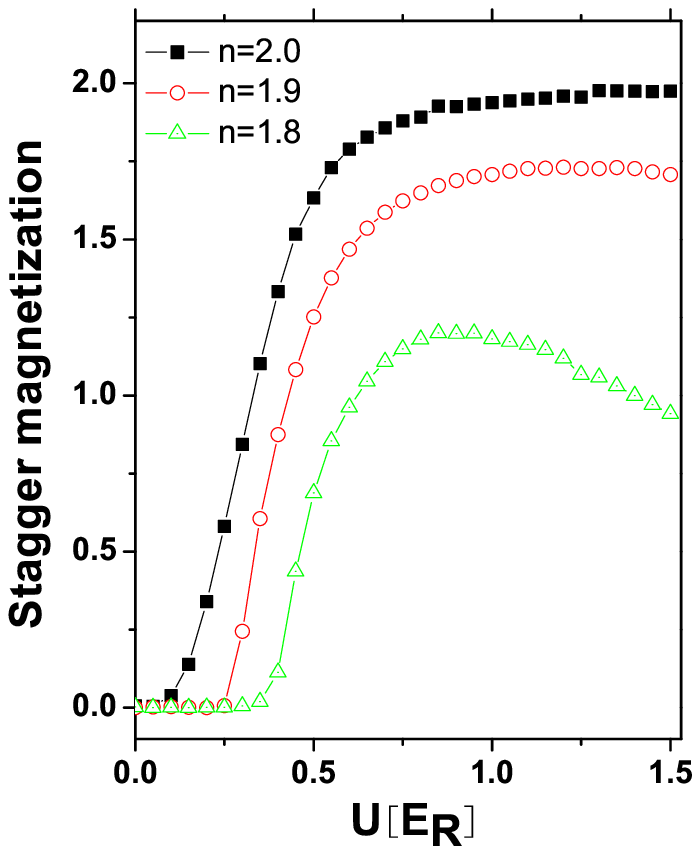}
\caption{\label{brandmag} (a). Quasi-particle weight v.s.
interaction strength $U$. $U^{\prime}$ and $J$ are varied with $U$
by keeping $U^\prime=J=0.299U$. At quarter filling (Red hollow dot)
the quasi-particle weight saturate to nonzero value, indicating an
absence of the Brinkman-Rice transition. (b). Stagger magnetization
v.s. interaction strength for several different fillings. At half
filling ($n=2$) the system crossover from weak coupling SDW behavior
to strong coupling spin-1 Neel state. Away from half filling, it's
needed a finite interaction strength to induce stagger
magnetization. }
\end{figure}

% $U$ is plotted in unit of the bare
%kinetic energy per site $\varepsilon_0=-0.273E_R$.

{\it Gutzwiller variation results:} For this complex two-band
Hubbard model we adopt a generalized version of the Gutzwiller
variational approach
\cite{gutzwiller}\cite{gebhard}\cite{daixi}\cite{kotliar97}\cite{LDA+G},
%Kotliar-Ruckenstein slave boson method,
Historically, the variational approach has been successfully used to
treat strong correlated systems such as normal $^3$He
\cite{vollhardt} and Mott transition \cite{BRtransition}.
Generalization of it is used to investigate physics associate with
band degeneracy, such as orbital selective Mott transition
\cite{daixi}\cite{koga} and transition metal ferromagnetism
\cite{kotliar97}. The variational wave function is
$|\Phi_G>=P_G|0>$, where $|0>$ is uncorrelated state which could be
fermi liquid, spin density wave or superconductivity ground state.
And $P_G=\prod_i \sum_{\Gamma} \lambda_{i,\Gamma}m_{i,\Gamma}$,
$m_{i,\Gamma}$ are projection operator onto atomic configurations,
$\lambda_{i, \Gamma}$ are corresponding variational parameters.
%The pseudo-fermion and slave boson
%fields satisfied several equations to guarantee the matrix elements
%in the physical space is identical with the original Hamiltonian.
%Treat the constrain on average, the saddle point solution of the
%action is proofed to be equivalent to the Gutzwiller approximation.
Under Gutzwiller approximation (GA)
\cite{gutzwiller}\cite{vollhardt} one can evaluate expected values
of operators over the projected wave function. By minimizing the
total energy one gets the ground state configurations, from which
one can identify various orders {\it e.g.}, charge density wave
order, orbital order, FM and AF order {\it etc}. GA neglects spatial
correlations and is only exact in infinite dimension, this
approximation has also been proven to be equivalent to the saddle
point of Kotliar-Ruckenstein slave boson functional
treatment\cite{kotliar}. Since it is non-perturbative in nature, the
variational approach treats the Fermi liquid as well as Mott
localized state on equal footing and hence gives a coherent
description of the intermediate coupling strength, connecting the
weak coupling mean field to the strong coupling perturbation
results\cite{zhai}. And the result of the present method is superior
to Hartree-Fock mean field that often overestimates the tendency
towards the ordered phases.

Disregard any long range order, at commensurate filling a reasonable
large on-site repulsive interaction would localize fermions and
drive the system into the paramagnetic Mott insulator phase. The
transition (Brinkman-Rice transition) could be described naturally
within Gutzwiller variational approach \cite{BRtransition} via
quasi-particle weight $Z$, which contributes to the coherent peak in
the spectral function. It could also be interpreted as the
renormalization factor of kinetic energy, the height of Fermi step
or inverse of mass enhancement factor. As $Z$ approaches to zero,
fermions becomes more and more localized and finally transforms into
the Mott insulator phase. However in the present case the large
Hund's rule coupling $J=U^{\prime}$ has dramatic effect on the
paramagnetic Mott transition: there is no Brinkman-Rice transition
at quarter filling, see Fig.\ref{brandmag}(a). The on-site
interaction Hamiltonian Eq.\ref{onsite} has doubly occupied
spin-triplet as its lowest energy states, which are degenerate with
the empty and singly occupied states. Charge fluctuation through
this channel is allowed at quarter filling thus the system does not
become localized when $U$ increases.
%could also be
%understood by dialogize of the on-site Hamiltonian shows that the
%due to Hund's rule coupling the low energy space contains spin
%triplet pairs of each orbital.
At half filling this kind of fluctuation is blocked by the large
on-site intra-orbital repulsive interaction $U$ and the
Brinkman-Rice transition manifests itself as shown in
Fig.\ref{brandmag}(a). However because of the huge spin entropy of
the paramagnetic Mott state, it would not act as the ground state
at zero temperature. When the system cools down the degeneracy
will be lifted and various magnetic/orbital orders develop
depending on the residual interactions between the spin/orbital
degrees of freedom.

Next we discuss the magnetic order of the p-band fermions. As the
previous section indicated, the band structure (Fermi surface
nesting and the Van Hove singularity) favors AF and FM order at
different energy scales, and the on-site Hund's rule coupling may
help stabilizing them. Thus, it is possible that these two phases
may exist even for weak or intermediately coupling strength. And it
is interesting to observe the coexistence of them in the optical
lattices with an external harmonic trap.

\begin{figure}
\includegraphics[width=7cm, height=5cm, bb=25 20 303 235]{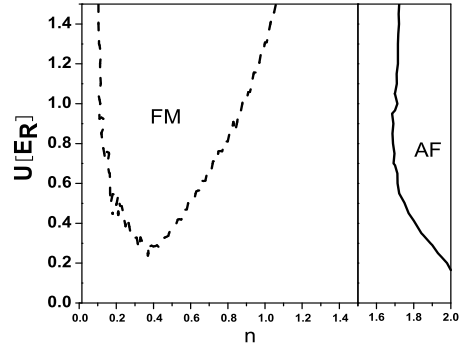}
\caption{\label{phasediag} Ground state phase diagram. Region
enclosed by full and dashed line is antiferromagnetic and
ferromagnetic phases respectively, the remainder part is in
paramagnetism phase. Contrary to the single band case, there is
robust ferromagnetic phase for vast range of filling and coupling
strength.}
\end{figure}

Fig. \ref{brandmag}(b) shows the development of the stagger
magnetization with increase of $U$ near half filling ($n=2$). Due to
the Fermi surface nesting, an arbitrary small $U$ drives the
instability of the two-sublattice antiferromagnetism at exact half
filing. At weak coupling it follows the spin density wave mean field
prediction, the stagger magnetization increases as $te^{-t/U}$.
Approaching the strong coupling limit the stagger magnetization
comes to its saturation value $2$. In this limit, the large
repulsive interaction quenches the charge degree of freedom while
the on-site Hund's rule coupling locks the local spins to form
spin-1 moments. Virtual hopping process leads to the
antiferromagnetic coupling of these local moments and the original
fermions model reduces to a spin-1 antiferromagnetic Heisenberg
model. This high spin AF order is more stable against the quantum
fluctuations. The present method also gives a coherent description
of the weak to strong crossover region, where the stagger
magnetization shows non-monotonic behavior for $n=1.8$ as shown in
Fig.\ref{brandmag}(b). The condensation energy of AF order is
roughly $W^2/U$, which is $10$ times larger than the $s$-band, see
Fig.\ref{pdos}(b)(c). Due to the large energy gain of the
antiferromagnetic order, we anticipate the transition temperature to
be two orders higher than that of the s-band case at the strong
coupling limit. At intermediate coupling strength the transition
temperature may attain its maximum value, which is accessible in
current cold atomic experiments.

Upon doping, the AF order is destroyed by movement of holes and the
system has featureless PM ground state. But at even lower fillings
where the Van Hove singularity plays a role the system has FM ground
state. Actually for the present depth of optical lattice, the
singularity in DOS plays the main role and the Hund's rule coupling
further stabilizes the tendency and enlarges the region of FM order
in the
phase diagram. %We turn off the Hund's rule coupling in our numerical
%calculations, {\it i.e.} $J=0$, $U^{\prime}=U$ for both equal and
%unequal spins, there is still finite ferromagnetism region in the
%phase diagram. Thus, we conclude that the dimensional reduction from
%the intrinsic anisotropy of the p-band is the main driving force of
%the robust ferromagnetism phase in the present case.
It is possible to tune the optical lattice to change the relative
weight of contribution of singularity in DOS and the Hund's rule
coupling, and this will shed light on the long controversial issue
of the mechanics of ferromagnetism in the transition metal oxides.
Due to the approximately particle-hole symmetry of the DOS, the
phase diagram shows a mirror symmetry with the $n=2$ axis, {\it
i.e.}, there is a FM region of similar shape at higher fillings.

\begin{figure}
\includegraphics[width=6.5cm, height=5.5cm,  bb=12 30 304 227]{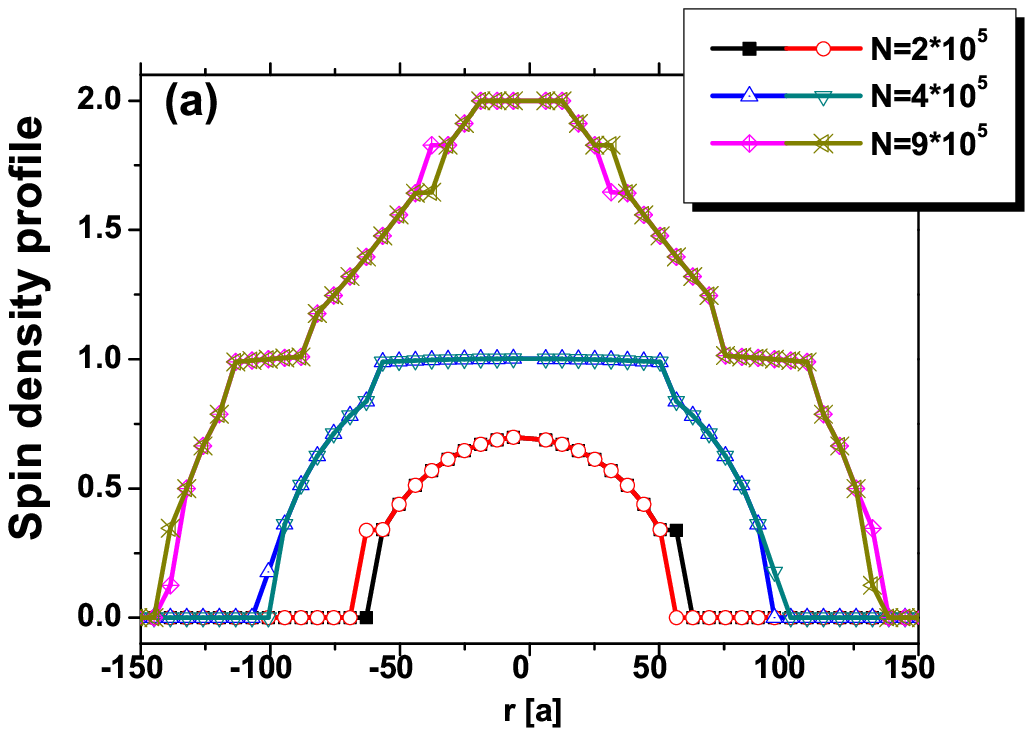}
\includegraphics[width=2cm, height=5.5cm,  bb=45 15 174 316]{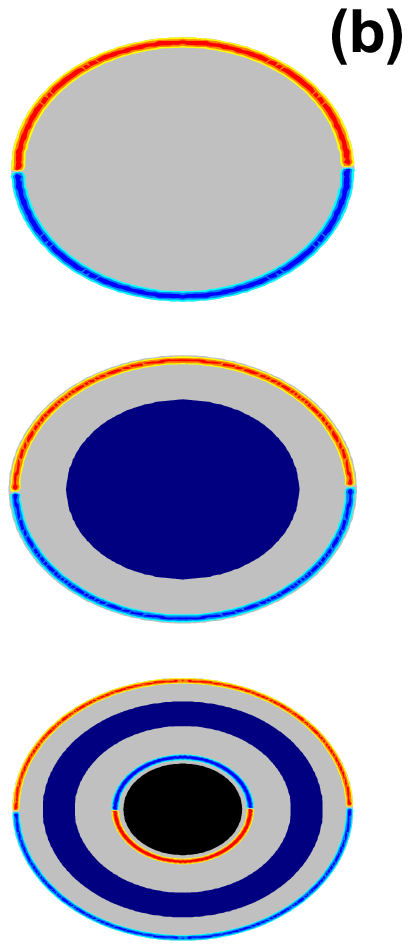}
\caption{\label{profile} (a). Side view of spin density profile in a
harmonic trap with different total number of atoms, two spin state
are $50:50$ mixture. The frequency of the external trap is
$\Omega=0.005E_R$. (b). Top view of various phases. Dark Blue region
denotes antiferromagntic phases, Black region denotes band insulator
and shallow blue/red for two kinds of polarized
fermions(ferromagnetic phase). Only density of atoms on the p-band
are shown in the figure, s-band spin densities provides a
homogeneous background near the center of the trap. }
\end{figure}

We now come to the effect of the external confinement. It is
provided by a combination of the magnetic trap and the Gaussian
profile of the laser beams. Combing with the local spin density
approximation(LSDA), our Gutzwiller variational approach provides a
natural way of incorporating spatial inhomogeneities into the
treatment of strong correlations. According to the celebrated
Hohenberg-Kohn theorem \cite{hktheorem}, the ground state energy of
an inhomogeneous system is solely determined by its ground state
spin density distribution. \eq E^{tot}= \int
d\textbf{r}\{E^{\textit{kin}+\textit{int}}[n_{\ua}(\textbf{r}),n_{\da}(\textbf{r})]+V^{\textit{ext}}(n_{\ua}(\textbf{r}),n_{\da}(\textbf{r}))\}\ee
where
$E^{\textit{kin}+\textit{int}}[n_{\ua}(\textbf{r}),n_{\da}(\textbf{r})]$
includes the kinetic energy, the lattice potential and the
interaction energy of fermions on an optical lattice, which depends
on the spin density distribution non-locally in general.
$V^{\textit{ext}}(n_{\ua}(\textbf{r}),n_{\da}(\textbf{r}))=\frac{1}{2}\Omega^2r^2(n_{\ua}(\textbf{r})+n_{\da}(\textbf{r}))$
is the trap energy. Usually $\Omega$ is much smaller compare to the
characteristic frequency of the optical lattice, thus the trap
provides a spatially slowly varying chemical potential. We could
expand the functional to the lowest order:
$E^{\textit{kin}+\textit{int}}[n_{\ua}(\textbf{r}),n_{\da}(\textbf{r})]\approx
E^{\textit{kin}+\textit{int}}(n_{\ua}(\textbf{r}),n_{\da}(\textbf{r}))$,
where the functional dependence on local spin densities could be
obtained by previously Gutzwiller variational calculation. Physical
picture of this strategy (LSDA) is to divide the whole system into
mesoscopic clusters of size
$l_{\Omega}=\sqrt{\frac{\hbar}{M\Omega}}$. Each of them feels
homogeneous external potential and at the same time could be treated
as in the thermodynamic limit, and Gutzwiller variational study
gives the dependence of energy and local spin densities within each
cluster.

We perform optimization of $E^{tot}$ by varying over spin density
distributions for $U=0.6E_R$ ($U^\prime$ and $J$ are determined
correspondingly) and several total number of atoms, the results are
shown in Fig. \ref{profile}. Densities of both spins decrease from
the trap center continuously to zero at the trap edge because of the
external potential. Due to the orbital degeneracy, maximum filling
at the trap center is $2$ for each spin, forming band insulator
phases. Competition of the external potential and cohesive energy of
the ordered phase gives fine structures of spin density profiles.
Regions where spin density around $1$ are more likely to enter into
AF phase, as long as the energy gain from AF phase exceeds
correspondingly energy loss from density redistribution. Plateaus
formed by AF (or PM Mott Insulator) phase was also reported in
previous Quantum Monte Carlo \cite{scalettar} and Dynamical Mean
Field theory \cite{dmft_af} study on single band inhomogeneous cold
fermionic atom systems. For regions with average filling less than
$0.5$ or larger than $1.5$ there are tendency towards ferromagnetic
order. But due to the conservation of total spins, FM region will
consist of two domains of half-shell shape with opposite spin
polarization, see Fig.\ref{profile}(a), similar structure is
anticipated in \cite{nagaokafm} concerns Nagaoka ferromagnetism. To
sum up, p-band cold fermionic atoms will form shell structures with
an external trap potential, for different radii the system crosses
the $U-n$ phase diagram and shows an band insulator, PM, FM and AF
phases Fig. \ref{profile}(b).

{\it Experimental signatures:} Population of higher bands have been
detected by time of flight (TOF) images
\cite{Kohl}\cite{highbandexpi}. Noise correlation
\cite{altman}\cite{blochnoise1} from TOF images may also detect spin
order through spin-spin correlation functions.
%By ramping down potentials
%adiabatically the crystal momentum is mapped to free particle
%momentum, the TOF images reflect the shape of the Fermi surface in
%extended Brillouin zone scheme.
Specifically, AF phase opens a charge transfer gap which could be
detected by Raman spectroscopy \cite{raman}, while the doubling of
unit cell might be detected by spin selective Bragg spectroscopy
\cite{spinBragg}. Spatial distribution of spin densities in harmonic
trap reported in this paper could be detected by spatially microwave
transition and spin-changing collisions techniques, which measure
the integrated density profiles along chosen directions
\cite{shellobser}.

{\it Summary:} Combining the band structure calculations and the
Gutzwliller variational approach, we perform a first principle
calculation of the zero temperature phase diagram of spin-$1/2$ cold
fermionic atoms on the two-fold degenerate p-band in an optical
lattice. We show that the system has robust ferromagnetic and
antiferromagnetic ground state at different fillings. We have traced
back the physical picture to the single particle feature included
anisotropy of orbital orientations and Fermi surface nesting as well
as correlated effects such as Hund' rule coupling. We also discuss
the inhomogeneous spatial distribution induced by an external
harmonic potential.

Acknowledgement: The work is supported by NSFC. Xie is supported
by US-DOE and NSF.

\end{document}